# 基於同態加密的相似度計算方法

Abel C. H. Chen
Chunghwa Telecom Co., Ltd.
chchen.scholar@gmail.com; ORCID 0000-0003-3628-3033

**摘要**

近年來，雖然有同態加密演算法被提出，可以提供加法同態加密、乘法同態加密。然而，為了在密文的情況下，提供查詢功能，仍需結合相似度計算。有鑑於此，本研究考慮餘弦相似度(cosine similarity)、角度相似度(angular similarity)、Tanimoto 相似度(Tanimoto similarity)、以及軟餘弦相似度(soft cosine similarity)等計算方法結合同態加密演算法。本研究提出數學模型證明，並且提供實例說明。在實驗結果中，實作不同安全強度下，證明提出的基於同態加密的各種相似度計算方法的效能。

**關鍵字**：同態加密、相似度、密碼學。

## 1. 前言

近年來，網路攻擊層出不窮，在駭客的攻擊下可能竊取到企業資料和個人隱私資料。因此，許多企業和個人也開始重視網路安全和雲端運算資料儲存安全。因此，開始有同態加密(Homomorphic Encryption)演算法[1]被提出，可以讓使用者在雲端環境儲存資料密文，從而避免資料明文在網路上傳輸和儲存，提升安全性。

其中，同態加密演算法可以讓資料在密文的情況下進行加法和乘法計算，並在計算完成後進行解密可以得到與明文進行加法和乘法計算一樣的結果。因此，需要在加密演算法和解密演算法上進行設計來達到上述功能。在應用上，同態加密演算法可以讓使用者在雲端環境只儲存資料密文，後續可以在密文狀態下進行使用者所需的加法和乘法計算。然而，為了在同態加密演算法的基礎上做到查詢功能，則需要再結合相似度(similarity)計算方法。因此，探索如何將同態加密演算法結合到各種相似度計算方法是重要的研究議題之一。

有鑑於此，本研究考慮餘弦相似度(cosine similarity)、角度相似度(angular similarity)、Tanimoto 相似度(Tanimoto similarity)、以及軟餘弦相似度(soft cosine similarity)等計算方法結合同態加密演算法。在第 3 節中提出數學模型證明，並且提供實例說明。

本論文共分為五個章節。第 2 節中介紹 ElGamal 同態加解密方法，以及說明各種常見的相似度計算方法。第 3 節提出本研究的基於同態加密的相似度計算方法，並且運用數學模型證明同態加密應用於各種相似度計算方法，並且提供實例說明。第 4 節參考美國國家標準與技術局(National Institute of Standards and Technology, NIST)，採用不同的安全強度(Security Strength)等級進行效能比較。第 5 節討論本研究貢獻，並且整理未來可行的研究方向。

## 2. 相關研究

本節將分別介紹 ElGamal 加解密方法[2]和相似度計算方法[3]。

### 2.1 ElGamal 加解密方法

本節將介紹 ElGamal 加解密方法，分別從金鑰產製方法、資料加密方法、以及資料解密方法來說明。

#### 2.1.1 金鑰產製方法

產製金鑰的過程中，將先隨機選擇一個足夠大的質數 $p$，存在一個整數模 $p$ 乘法群 $Z_p = \{1, ...,$

$p$ - 1}，並有生成元 $g$。再隨機選擇一個小於 $p$ - 1 的整數作為私鑰 $q$，以及運用公式(1)計算公鑰 $Q$。其中，質數 $p$、生成元 $g$、以及公鑰 $Q$ 為可公開。

$$Q = g^q \pmod{p} \tag{1}$$

### 2.1.2 資料加密方法

資料加密方法對明文 $x$ 加密的過程中，可產生一隨機數 $r$ 作為加密使用，如公式(2)和公式(3)所示。其中，加密後的密文為 $(c_1, c_2)$。

$$c_1 = g^r \pmod{p} \tag{2}$$

$$c_2 = xR, \text{ where } R = Q^r \pmod{p} \tag{3}$$

### 2.1.3 資料解密方法

資料解密方法可運用公式(4)以私鑰 $q$ 對密文 $(c_1, c_2)$ 解密取得明文 $x$。

$$\begin{aligned}
[c_2(c_1{}^q)^{-1}] \bmod p &= \left(\frac{c_2}{c_1{}^q}\right) \pmod{p} \\
&= \left(\frac{xR}{g^{rq}}\right) \pmod{p} \\
&= \left(\frac{xQ^r}{g^{rq}}\right) \pmod{p} \\
&= \left(\frac{xg^{qr}}{g^{rq}}\right) \pmod{p} = x
\end{aligned} \tag{4}$$

## 2.2 相似度計算方法

本節將介紹相似度計算方法，包含餘弦相似度、角度相似度、Tanimoto 相似度、以及軟餘弦相似度。

### 2.2.1 餘弦相似度

餘弦相似度可以用來計算向量之間夾角的餘弦值作為向量間的相似度。例如，具有 $n$ 維的向量 $\mathbf{X}_i = \{x_{i,1}, x_{i,2}, …, x_{i,n}\}$ 和向量 $\mathbf{X}_j = \{x_{j,1}, x_{j,2}, …, x_{j,n}\}$ 的餘弦相似度 $s_{c,i,j}$ 可以通過公式(5)計算。並且，可以通過公式(6)取得餘弦距離 $d_{c,i,j}$。

$$\begin{aligned}
s_{c,i,j} &= \frac{\mathbf{X}_i \cdot \mathbf{X}_j}{\|\mathbf{X}_i\|\|\mathbf{X}_j\|} \\
&= \frac{\sum_{k=1}^{n} x_{i,k} x_{j,k}}{\sqrt{\sum_{k=1}^{n} x_{i,k}{}^2} \sqrt{\sum_{k=1}^{n} x_{j,k}{}^2}}
\end{aligned} \tag{5}$$

$$\begin{aligned}
d_{c,i,j} &= 1 - \frac{\mathbf{X}_i \cdot \mathbf{X}_j}{\|\mathbf{X}_i\|\|\mathbf{X}_j\|} \\
&= 1 - \frac{\sum_{k=1}^{n} x_{i,k} x_{j,k}}{\sqrt{\sum_{k=1}^{n} x_{i,k}{}^2} \sqrt{\sum_{k=1}^{n} x_{j,k}{}^2}}
\end{aligned} \tag{6}$$

### 2.2.2 角度相似度

角度相似度類似於餘弦相似度，可以通過向量之間夾角的角度作為向量間的相似度。例如，具有 $n$ 維的向量 $\mathbf{X}_i$ 和向量 $\mathbf{X}_j$ 的角度相似度 $s_{a,i,j}$ 可以通過公式(7)計算。並且，可以通過公式(8)取得角度距離 $d_{a,i,j}$。

### 2.2.3 Tanimoto 相似度

Tanimoto 相似度類似於 Jaccard 相似度，以兩集合的交集作為分子，並以兩集合的聯集作為分母，而且 Tanimoto 相似度計算可推廣至兩個向量間的相似度計算。例如，具有 $n$ 維的向量 $\mathbf{X}_i$ 和向量 $\mathbf{X}_j$ 的 Tanimoto 相似度 $s_{t,i,j}$ 可以通過公式(9)計算。並且，可以通過公式(10)取得 Tanimoto 距離 $d_{t,i,j}$，從公式(10)可以觀察出分子為兩向量的各分量相減

$$\begin{aligned}
s_{a,i,j} &= 1 - \frac{2 \times \arccos(s_{c,i,j})}{\pi} \\
&= 1 - \frac{2 \times \arccos\left(\frac{\mathbf{X}_i \cdot \mathbf{X}_j}{\|\mathbf{X}_i\|\|\mathbf{X}_j\|}\right)}{\pi} \\
&= 1 - \frac{2 \times \arccos\left(\frac{\sum_{k=1}^{n} x_{i,k} x_{j,k}}{\sqrt{\sum_{k=1}^{n} x_{i,k}{}^2} \sqrt{\sum_{k=1}^{n} x_{j,k}{}^2}}\right)}{\pi}
\end{aligned} \tag{7}$$

的平方和作為距離。

### 2.2.4　軟餘弦相似度

軟餘弦相似度相較於餘弦相似度還考慮了特徵間的相似度進行加權。例如，具有 $n$ 維的向量 $\mathbf{X}_i$ 和向量 $\mathbf{X}_j$ 的軟餘弦相似度 $s_{s,i,j}$ 可以通過公式(11)計算。其中，第 $k$ 個特徵與第 $l$ 個特徵的相似度是 $w_{k,l}$。並且，可以通過公式(8)取得軟餘弦距離 $d_{s,i,j}$。

## 3. 研究方法

本研究先在 3.1 節定義同態加解密演算法的目標，3.2 節在符合同態加解密演算法下設計各種相似度計算方法，並且通過數學模型進行證明。3.3 節中提供實例來說明基於同態加密的各種相似度計算方法。

### 3.1　同態加解密演算法的目標

為定義同態加解密演算法的目標，本節假設明文為 $x_i$，通過加密函數 $Enc(x_i)$ 計算後可以得到密文 $c_{x_i}$，如公式(13)所示；並且有解密函數 $Dec(c_{x_i})$ 可對密文 $c_{x_i}$ 進行解密後得到 $x_i$，如公式(14)所示。本研究中以乘法同態加解密演算法為例，假設密文 $c_{x_i}$ 相乘後得到 $\Omega$，如公式(15)所示；而明文 $x_i$ 相乘後得到 $\Psi$，並且 $\Psi$ 加密後與密文相乘結果一樣為 $\Omega$，如

$$
\begin{aligned}
d_{a,i,j} &= \frac{2 \times \arccos(\alpha_{i,j})}{\pi} \\
&= \frac{2 \times \arccos\left(\dfrac{\mathbf{X}_i \cdot \mathbf{X}_j}{\|\mathbf{X}_i\|\|\mathbf{X}_j\|}\right)}{\pi} \\
&= \frac{2 \times \arccos\left(\dfrac{\sum_{k=1}^n x_{i,k} x_{j,k}}{\sqrt{\sum_{k=1}^n x_{i,k}{}^2}\sqrt{\sum_{k=1}^n x_{j,k}{}^2}}\right)}{\pi}
\end{aligned} \tag{8}
$$

$$
\begin{aligned}
s_{t,i,j} &= \frac{\mathbf{X}_i \cdot \mathbf{X}_j}{\|\mathbf{X}_i\|^2 + \|\mathbf{X}_j\|^2 - \mathbf{X}_i \cdot \mathbf{X}_j} \\
&= \frac{\sum_{k=1}^n x_{i,k} x_{j,k}}{\left(\sqrt{\sum_{k=1}^n x_{i,k}{}^2}\right)^2 + \left(\sqrt{\sum_{k=1}^n x_{j,k}{}^2}\right)^2 - \sum_{k=1}^n x_{i,k} x_{j,k}}
\end{aligned} \tag{9}
$$

$$
\begin{aligned}
d_{t,i,j} &= 1 - \frac{\mathbf{X}_i \cdot \mathbf{X}_j}{\|\mathbf{X}_i\|^2 + \|\mathbf{X}_j\|^2 - \mathbf{X}_i \cdot \mathbf{X}_j} \\
&= \frac{\|\mathbf{X}_i\|^2 + \|\mathbf{X}_j\|^2 - 2 \times \mathbf{X}_i \cdot \mathbf{X}_j}{\|\mathbf{X}_i\|^2 + \|\mathbf{X}_j\|^2 - \mathbf{X}_i \cdot \mathbf{X}_j} \\
&= \frac{\|\mathbf{X}_i - \mathbf{X}_j\|^2}{\|\mathbf{X}_i\|^2 + \|\mathbf{X}_j\|^2 - \mathbf{X}_i \cdot \mathbf{X}_j} \\
&= \frac{\sum_{k=1}^n (x_{i,k} - x_{j,k})^2}{\left(\sqrt{\sum_{k=1}^n x_{i,k}{}^2}\right)^2 + \left(\sqrt{\sum_{k=1}^n x_{j,k}{}^2}\right)^2 - \sum_{k=1}^n x_{i,k} x_{j,k}}
\end{aligned} \tag{10}
$$

$$
s_{s,i,j} = \frac{\sum_{k=1}^n \sum_{l=1}^n w_{k,l} x_{i,k} x_{j,l}}{\sqrt{\sum_{k=1}^n \sum_{l=1}^n w_{k,l} x_{i,k} x_{i,l}} \sqrt{\sum_{k=1}^n \sum_{l=1}^n w_{k,l} x_{j,k} x_{j,l}}} \tag{11}
$$

$$
d_{s,i,j} = 1 - \frac{\sum_{k=1}^n \sum_{l=1}^n w_{k,l} x_{i,k} x_{j,l}}{\sqrt{\sum_{k=1}^n \sum_{l=1}^n w_{k,l} x_{i,k} x_{i,l}} \sqrt{\sum_{k=1}^n \sum_{l=1}^n w_{k,l} x_{j,k} x_{j,l}}} \tag{12}
$$

公式(16)所示。因此，對密文$c_{x_i}$相乘後結果Ω解密可得Ψ，如公式(17)所示。因此，建構符合公式(13)~(17)的加密函數$Enc(x_i)$和解密函數$Dec(c_{x_i})$，即符合同態加解密演算法的目標。

$$c_{x_i} = Enc(x_i) \tag{13}$$

$$x_i = Dec(c_{x_i}) \tag{14}$$

$$\Omega = \prod_{i=1}^{n} c_{x_i} \tag{15}$$

$$\Omega = Enc(\Psi), \text{where } \Psi = \prod_{i=1}^{n} x_i \tag{16}$$

$$\Psi = Dec(\Omega) = Dec\left(\prod_{i=1}^{n} c_{x_i}\right) = \prod_{i=1}^{n} x_i \tag{17}$$

本研究採用 ElGamal 加解密方法作為同態加解密演算法，數學證明如公式(18)~(20)所示。假設有質數 $p$、生成元 $g$、以及公鑰 $Q$(如公式(1)所示)。在加密過程中，為簡化說明，採用相同隨機數 $r$ 作為加密使用，可得 $c_1$(如公式(2)所示)，加密計算如公式(18)所示；然而，為提升安全性可採用不同的隨機數，計算結果會是一致的。單一筆密文$c_{x_i}$解密計算如公式(19)所示，並且通過公式(20)證明符合同態加解密演算法的目標(即符合公式(15)~(17))。

$$Enc(x_i) = x_i R, \text{where } R = Q^r \pmod{p}$$
$$= c_{x_i} \tag{18}$$

$$Dec(c_{x_i}) = [c_{x_i}(c_1{}^q)^{-1}] \pmod{p}$$
$$= \left(\frac{c_{x_i}}{c_1{}^q}\right) \pmod{p}$$
$$= \left(\frac{x_i R}{g^{rq}}\right) \pmod{p} \tag{19}$$
$$= \left(\frac{x_i Q^r}{g^{rq}}\right) \pmod{p}$$
$$= \left(\frac{x_i g^{qr}}{g^{rq}}\right) \pmod{p} = x_i$$

$$Dec(\Omega) = \left[\prod_{i=1}^{n} c_{x_i}(c_1{}^q)^{-1}\right] \pmod{p}$$
$$= \left(\prod_{i=1}^{n}\left(\frac{c_{x_i}}{c_1{}^q}\right) \pmod{p}\right)$$
$$= \left(\prod_{i=1}^{n}\left(\frac{x_i R}{g^{rq}}\right) \pmod{p}\right)$$
$$= \left(\prod_{i=1}^{n}\left(\frac{x_i Q^r}{g^{rq}}\right) \pmod{p}\right) \tag{20}$$
$$= \left(\prod_{i=1}^{n}\left(\frac{x_i g^{qr}}{g^{rq}}\right) \pmod{p}\right)$$
$$= \prod_{i=1}^{n} x_i = \Psi$$

### 3.2 基於同態加密的相似度計算方法及數學證明

本節將 ElGamal 加解密方法作為同態加解密演算法結合到各種相似度計算方法，包含有基於同態加密的餘弦相似度、基於同態加密的角度相似度、基於同態加密的 Tanimoto 相似度、以及基於同態加密的軟餘弦相似度，分別於 3.2.1 節~3.2.4 節中證明。

為證明基於同態加密的相似度計算方法，本研究對向量明文 $\mathbf{X}_i = \{x_{i,1}, x_{i,2}, …, x_{i,n}\}$ 以加密函數 $Enc(x_{i,k})$ 加密後可得向量密文 $\mathbf{C}_i = \{c_{i,1}, c_{i,2}, …, c_{i,n}\}$，如公式(21)所示。並且，可以通過解密函數 $Dec(c_{i,k})$ 將密文 $c_{i,k}$ 解密為明文 $x_{i,k}$，如公式(22)所示。

$$c_{i,k} = Enc(x_{i,k}) \tag{21}$$

$$x_{i,k} = Dec(c_{i,k}) \tag{22}$$

#### 3.2.1 基於同態加密的餘弦相似度

為證明基於同態加密的餘弦相似度計算方法，

本節採用具有 $n$ 維的向量密文 $\mathbf{C}_i = \{c_{i,1}, c_{i,2}, ..., c_{i,n}\}$ 和向量密文 $\mathbf{C}_j = \{c_{j,1}, c_{j,2}, ..., c_{j,n}\}$，計算對應的向量 $\mathbf{X}_i$ 和向量 $\mathbf{X}_j$ 的餘弦相似度 $s_{c,i,j}$，如公式(23)所示。並且，可以通過公式(24)取得餘弦距離 $d_{c,i,j}$。

本節採用具有 $n$ 維的向量密文 $\mathbf{C}_i$ 和向量密文 $\mathbf{C}_j$，基於公式(23)提出公式(25)證明向量密文 $\mathbf{C}_i$ 和向量密文 $\mathbf{C}_j$ 的角度相似度解密後與向量 $\mathbf{X}_i$ 和向量 $\mathbf{X}_j$ 的角度相似度 $s_{a,i,j}$ 相同。並且，可以通過公式(26)取得角度距離 $d_{a,i,j}$。

### 3.2.2 基於同態加密的角度相似度

為證明基於同態加密的角度相似度計算方法，

$$\begin{aligned}
\frac{Dec(\mathbf{C}_i \cdot \mathbf{C}_j)}{Dec(\|\mathbf{C}_i\|\|\mathbf{C}_j\|)} &= \frac{\left[\sum_{k=1}^{n} c_{i,k}(c_1{}^q)^{-1} c_{j,k}(c_1{}^q)^{-1}\right] (\bmod\ p)}{\left[\sqrt{\sum_{k=1}^{n}\left(c_{i,k}(c_1{}^q)^{-1}\right)^2} \sqrt{\sum_{k=1}^{n}\left(c_{j,k}(c_1{}^q)^{-1}\right)^2}\right] (\bmod\ p)} \\
&= \frac{\left[\sum_{k=1}^{n}\left(\frac{c_{i,k}}{c_1{}^q}\right)\left(\frac{c_{j,k}}{c_1{}^q}\right)\right] (\bmod\ p)}{\left[\sqrt{\sum_{k=1}^{n}\left(\frac{c_{i,k}}{c_1{}^q}\right)^2} \sqrt{\sum_{k=1}^{n}\left(\frac{c_{j,k}}{c_1{}^q}\right)^2}\right] (\bmod\ p)} \\
&= \frac{\left[\sum_{k=1}^{n}\left(\frac{x_{i,k}R}{g^{rq}}\right)\left(\frac{x_{j,k}R}{g^{rq}}\right)\right] (\bmod\ p)}{\left[\sqrt{\sum_{k=1}^{n}\left(\frac{x_{i,k}R}{g^{rq}}\right)^2} \sqrt{\sum_{k=1}^{n}\left(\frac{x_{j,k}R}{g^{rq}}\right)^2}\right] (\bmod\ p)} \\
&= \frac{\left[\sum_{k=1}^{n}\left(\frac{x_{i,k}Q^r}{g^{rq}}\right)\left(\frac{x_{j,k}Q^r}{g^{rq}}\right)\right] (\bmod\ p)}{\left[\sqrt{\sum_{k=1}^{n}\left(\frac{x_{i,k}Q^r}{g^{rq}}\right)^2} \sqrt{\sum_{k=1}^{n}\left(\frac{x_{j,k}Q^r}{g^{rq}}\right)^2}\right] (\bmod\ p)} \\
&= \frac{\left[\sum_{k=1}^{n}\left(\frac{x_{i,k}g^{qr}}{g^{rq}}\right)\left(\frac{x_{j,k}g^{qr}}{g^{rq}}\right)\right] (\bmod\ p)}{\left[\sqrt{\sum_{k=1}^{n}\left(\frac{x_{i,k}g^{qr}}{g^{rq}}\right)^2} \sqrt{\sum_{k=1}^{n}\left(\frac{x_{j,k}g^{qr}}{g^{rq}}\right)^2}\right] (\bmod\ p)} \\
&= \frac{\sum_{k=1}^{n} x_{i,k} x_{j,k}}{\sqrt{\sum_{k=1}^{n} x_{i,k}{}^2} \sqrt{\sum_{k=1}^{n} x_{j,k}{}^2}} = \frac{\mathbf{X}_i \cdot \mathbf{X}_j}{\|\mathbf{X}_i\|\|\mathbf{X}_j\|} = s_{c,i,j}
\end{aligned} \tag{23}$$

$$1 - \frac{Dec(\mathbf{C}_i \cdot \mathbf{C}_j)}{Dec(\|\mathbf{C}_i\|\|\mathbf{C}_j\|)} = 1 - \frac{\mathbf{X}_i \cdot \mathbf{X}_j}{\|\mathbf{X}_i\|\|\mathbf{X}_j\|} = 1 - s_{c,i,j} = d_{c,i,j} \tag{24}$$

$$\begin{aligned}
1 - \frac{2 \times \arccos\left(\frac{Dec(\mathbf{C}_i \cdot \mathbf{C}_j)}{Dec(\|\mathbf{C}_i\|\|\mathbf{C}_j\|)}\right)}{\pi} &= 1 - \frac{2 \times \arccos\left(\frac{\mathbf{X}_i \cdot \mathbf{X}_j}{\|\mathbf{X}_i\|\|\mathbf{X}_j\|}\right)}{\pi} \\
&= 1 - \frac{2 \times \arccos(s_{c,i,j})}{\pi} = s_{a,i,j}
\end{aligned} \tag{25}$$

$$\frac{2 \times \arccos\left(\frac{Dec(\mathbf{C}_i \cdot \mathbf{C}_j)}{Dec(\|\mathbf{C}_i\|\|\mathbf{C}_j\|)}\right)}{\pi} = \frac{2 \times \arccos(s_{c,i,j})}{\pi} = d_{a,i,j} \tag{26}$$

$$\frac{Dec(\mathbf{C}_i \cdot \mathbf{C}_j)}{Dec(\|\mathbf{C}_i\|^2) + Dec(\|\mathbf{C}_j\|^2) - Dec(\mathbf{C}_i \cdot \mathbf{C}_j)}$$

$$= \frac{\left[\sum_{k=1}^{n}\left(\frac{c_{i,k}}{c_1{}^q}\right)\left(\frac{c_{j,k}}{c_1{}^q}\right)\right](\bmod\ p)}{\left[\left(\sum_{k=1}^{n}\left(\frac{c_{i,k}}{c_1{}^q}\right)^2\right)(\bmod\ p)\right] + \left[\left(\sum_{k=1}^{n}\left(\frac{c_{j,k}}{c_1{}^q}\right)^2\right)(\bmod\ p)\right] - \left[\sum_{k=1}^{n}\left(\frac{c_{i,k}}{c_1{}^q}\right)\left(\frac{c_{j,k}}{c_1{}^q}\right)\right](\bmod\ p)}$$

$$= \frac{\left[\sum_{k=1}^{n}\left(\frac{x_{i,k}R}{g^{rq}}\right)\left(\frac{x_{j,k}R}{g^{rq}}\right)\right](\bmod\ p)}{\left[\left(\sum_{k=1}^{n}\left(\frac{x_{i,k}R}{g^{rq}}\right)^2\right)(\bmod\ p)\right] + \left[\left(\sum_{k=1}^{n}\left(\frac{x_{j,k}R}{g^{rq}}\right)^2\right)(\bmod\ p)\right] - \left[\sum_{k=1}^{n}\left(\frac{x_{i,k}R}{g^{rq}}\right)\left(\frac{x_{j,k}R}{g^{rq}}\right)\right](\bmod\ p)}$$

$$= \frac{\left[\sum_{k=1}^{n}\left(\frac{x_{i,k}Q^r}{g^{rq}}\right)\left(\frac{x_{j,k}Q^r}{g^{rq}}\right)\right](\bmod\ p)}{\left[\left(\sum_{k=1}^{n}\left(\frac{x_{i,k}Q^r}{g^{rq}}\right)^2\right)(\bmod\ p)\right] + \left[\left(\sum_{k=1}^{n}\left(\frac{x_{j,k}Q^r}{g^{rq}}\right)^2\right)(\bmod\ p)\right] - \left[\sum_{k=1}^{n}\left(\frac{x_{i,k}Q^r}{g^{rq}}\right)\left(\frac{x_{j,k}Q^r}{g^{rq}}\right)\right](\bmod\ p)} \quad (27)$$

$$= \frac{\left[\sum_{k=1}^{n}\left(\frac{x_{i,k}g^{qr}}{g^{rq}}\right)\left(\frac{x_{j,k}g^{qr}}{g^{rq}}\right)\right](\bmod\ p)}{\left[\left(\sum_{k=1}^{n}\left(\frac{x_{i,k}g^{qr}}{g^{rq}}\right)^2\right)(\bmod\ p)\right] + \left[\left(\sum_{k=1}^{n}\frac{x_{j,k}g^{qr}{}^2}{g^{rq}}\right)(\bmod\ p)\right] - \left[\sum_{k=1}^{n}\left(\frac{x_{i,k}g^{qr}}{g^{rq}}\right)\left(\frac{x_{j,k}g^{qr}}{g^{rq}}\right)\right](\bmod\ p)}$$

$$= \frac{\sum_{k=1}^{n} x_{i,k} x_{j,k}}{\left(\sqrt{\sum_{k=1}^{n} x_{i,k}{}^2}\right)^2 + \left(\sqrt{\sum_{k=1}^{n} x_{j,k}{}^2}\right)^2 - \sum_{k=1}^{n} x_{i,k} x_{j,k}}$$

$$= \frac{\mathbf{X}_i \cdot \mathbf{X}_j}{\|\mathbf{X}_i\|^2 + \|\mathbf{X}_j\|^2 - \mathbf{X}_i \cdot \mathbf{X}_j} = s_{t,i,j}$$

$$1 - \frac{Dec(\mathbf{C}_i \cdot \mathbf{C}_j)}{Dec(\|\mathbf{C}_i\|^2) + Dec(\|\mathbf{C}_j\|^2) - Dec(\mathbf{C}_i \cdot \mathbf{C}_j)} = 1 - \frac{\mathbf{X}_i \cdot \mathbf{X}_j}{\|\mathbf{X}_i\|^2 + \|\mathbf{X}_j\|^2 - \mathbf{X}_i \cdot \mathbf{X}_j} = d_{t,i,j} \quad (28)$$

### 3.2.3 基於同態加密的 Tanimoto 相似度

為證明基於同態加密的 Tanimoto 相似度計算方法，本節採用具有 $n$ 維的向量密文 $\mathbf{C}_i$ 和向量密文 $\mathbf{C}_j$，提出公式(27)證明向量密文 $\mathbf{C}_i$ 和向量密文 $\mathbf{C}_j$ 的 Tanimoto 相似度解密後與向量 $\mathbf{X}_i$ 和向量 $\mathbf{X}_j$ 的 Tanimoto 相似度 $s_{t,i,j}$ 相同。其中，原始的 Tanimoto 相似度計算上包含有分子(即 $\mathbf{X}_i \cdot \mathbf{X}_j$)和分母(即 $\|\mathbf{X}_i\|^2$、$\|\mathbf{X}_j\|^2$、$\mathbf{X}_i \cdot \mathbf{X}_j$)，對每個部分各別做密文乘法和加總後進行解密。除此之外，可以在公式(27)的基礎上通過公式(28)計算 Tanimoto 距離，可證明與明文計算結果 $d_{t,i,j}$ 一致。

### 3.2.4 基於同態加密的軟餘弦相似度

為證明基於同態加密的軟餘弦相似度計算方法，本節假設第 $k$ 個特徵與第 $l$ 個特徵的相似度是 $w_{k,l}$，通過公式(29)加密後得到密文 $\omega_{k,l}$。通過 $\omega_{k,l}$ 進行特徵間的相似度加權。例如，具有 $n$ 維的向量密文 $\mathbf{C}_i$ 和向量密文 $\mathbf{C}_j$ 的軟餘弦相似度 $s_{s,i,j}$ 可以通過公式(30)證明。除此之外，可以在公式(30)的基礎上通過公式(31)計算軟餘弦距離，可證明與明文計算結果 $d_{s,i,j}$ 一致。

$$\omega_{k,l} = Enc(w_{k,l}) \quad (29)$$

### 3.3 實例說明

為說明本研究提出基於同態加密的各種相似度計算方法，本節提供計算實例進行說明。然而，為了簡單說明，本節採用的質數較小，未來部署到實際系統時可採用大質數提升安全性；在第 4 節將提供在不同安全強度下的計算，

本節採用下列參數值進行說明：質數 $p = 2{,}932{,}031{,}007{,}403$、生成元 $g = 3$、隨機數 $r = 2$、以及私鑰 $q = 5$。因此，通過公式(1)~(3)，分別可以計算出、以及公鑰 $Q = 3^5\ (\bmod\ p) = 243$、$c_1 =$

$$
\frac{Dec(\sum_{k=1}^{n}\sum_{l=1}^{n}\omega_{k,l}c_{i,k}c_{j,l})}{Dec(\sqrt{\sum_{k=1}^{n}\sum_{l=1}^{n}\omega_{k,l}c_{i,k}c_{i,l}}\sqrt{\sum_{k=1}^{n}\sum_{l=1}^{n}\omega_{k,l}c_{j,k}c_{j,l}})}
$$

$$
=\frac{\left[\sum_{k=1}^{n}\sum_{l=1}^{n}\left(\frac{\omega_{k,l}}{c_1{}^q}\right)\left(\frac{c_{i,k}}{c_1{}^q}\right)\left(\frac{c_{j,l}}{c_1{}^q}\right)\right](\bmod\ p)}{\left[\sqrt{\sum_{k=1}^{n}\sum_{l=1}^{n}\left(\frac{\omega_{k,l}}{c_1{}^q}\right)\left(\frac{c_{i,k}}{c_1{}^q}\right)\left(\frac{c_{i,l}}{c_1{}^q}\right)}\sqrt{\sum_{k=1}^{n}\sum_{l=1}^{n}\left(\frac{\omega_{k,l}}{c_1{}^q}\right)\left(\frac{c_{j,k}}{c_1{}^q}\right)\left(\frac{c_{j,l}}{c_1{}^q}\right)}\right](\bmod\ p)}
$$

$$
=\frac{\left[\sum_{k=1}^{n}\sum_{l=1}^{n}\left(\frac{w_{k,l}R}{g^{rq}}\right)\left(\frac{x_{i,k}R}{g^{rq}}\right)\left(\frac{x_{j,l}R}{g^{rq}}\right)\right](\bmod\ p)}{\left[\sqrt{\sum_{k=1}^{n}\sum_{l=1}^{n}\left(\frac{w_{k,l}R}{g^{rq}}\right)\left(\frac{x_{i,k}R}{g^{rq}}\right)\left(\frac{x_{i,l}R}{g^{rq}}\right)}\sqrt{\sum_{k=1}^{n}\sum_{l=1}^{n}\left(\frac{w_{k,l}R}{g^{rq}}\right)\left(\frac{x_{j,k}R}{g^{rq}}\right)\left(\frac{x_{j,l}R}{g^{rq}}\right)}\right](\bmod\ p)}
$$

$$
=\frac{\left[\sum_{k=1}^{n}\sum_{l=1}^{n}\left(\frac{w_{k,l}Q^r}{g^{rq}}\right)\left(\frac{x_{i,k}Q^r}{g^{rq}}\right)\left(\frac{x_{j,l}Q^r}{g^{rq}}\right)\right](\bmod\ p)}{\left[\sqrt{\sum_{k=1}^{n}\sum_{l=1}^{n}\left(\frac{w_{k,l}Q^r}{g^{rq}}\right)\left(\frac{x_{i,k}Q^r}{g^{rq}}\right)\left(\frac{x_{i,l}Q^r}{g^{rq}}\right)}\sqrt{\sum_{k=1}^{n}\sum_{l=1}^{n}\left(\frac{w_{k,l}Q^r}{g^{rq}}\right)\left(\frac{x_{j,k}Q^r}{g^{rq}}\right)\left(\frac{x_{j,l}Q^r}{g^{rq}}\right)}\right](\bmod\ p)} \quad (30)
$$

$$
=\frac{\left[\sum_{k=1}^{n}\sum_{l=1}^{n}\left(\frac{w_{k,l}Q^r}{g^{rq}}\right)\left(\frac{x_{i,k}Q^r}{g^{rq}}\right)\left(\frac{x_{j,l}Q^r}{g^{rq}}\right)\right](\bmod\ p)}{\left[\sqrt{\sum_{k=1}^{n}\sum_{l=1}^{n}\left(\frac{w_{k,l}Q^r}{g^{rq}}\right)\left(\frac{x_{i,k}Q^r}{g^{rq}}\right)\left(\frac{x_{i,l}Q^r}{g^{rq}}\right)}\sqrt{\sum_{k=1}^{n}\sum_{l=1}^{n}\left(\frac{w_{k,l}Q^r}{g^{rq}}\right)\left(\frac{x_{j,k}Q^r}{g^{rq}}\right)\left(\frac{x_{j,l}Q^r}{g^{rq}}\right)}\right](\bmod\ p)}
$$

$$
=\frac{\left[\sum_{k=1}^{n}\sum_{l=1}^{n}\left(\frac{w_{k,l}g^{qr}}{g^{rq}}\right)\left(\frac{x_{i,k}g^{qr}}{g^{rq}}\right)\left(\frac{x_{j,l}g^{qr}}{g^{rq}}\right)\right](\bmod\ p)}{\left[\sqrt{\sum_{k=1}^{n}\sum_{l=1}^{n}\left(\frac{w_{k,l}g^{qr}}{g^{rq}}\right)\left(\frac{x_{i,k}g^{qr}}{g^{rq}}\right)\left(\frac{x_{i,l}g^{qr}}{g^{rq}}\right)}\sqrt{\sum_{k=1}^{n}\sum_{l=1}^{n}\left(\frac{w_{k,l}g^{qr}}{g^{rq}}\right)\left(\frac{x_{j,k}g^{qr}}{g^{rq}}\right)\left(\frac{x_{j,l}g^{qr}}{g^{rq}}\right)}\right](\bmod\ p)}
$$

$$
=\frac{\sum_{k=1}^{n}\sum_{l=1}^{n}w_{k,l}x_{i,k}x_{j,l}}{\sqrt{\sum_{k=1}^{n}\sum_{l=1}^{n}w_{k,l}x_{i,k}x_{i,l}}\sqrt{\sum_{k=1}^{n}\sum_{l=1}^{n}w_{k,l}x_{j,k}x_{j,l}}}=s_{s,i,j}
$$

$$
1-\frac{Dec(\sum_{k=1}^{n}\sum_{l=1}^{n}\omega_{k,l}c_{i,k}c_{j,l})}{Dec(\sqrt{\sum_{k=1}^{n}\sum_{l=1}^{n}\omega_{k,l}c_{i,k}c_{i,l}}\sqrt{\sum_{k=1}^{n}\sum_{l=1}^{n}\omega_{k,l}c_{j,k}c_{j,l}})}=1-s_{s,i,j}=d_{s,i,j} \quad (31)
$$

$3^2\ (\bmod\ p) = 9$、以及 $R = 243^2\ (\bmod\ p) = 59049$，作為加密時使用。除此之外，假設資料維度 $n = 3$，兩個向量值分別是 $\mathbf{X}_i = \{x_{i,1}, x_{i,2}, x_{i,3}\} = \{1, 2, 3\}$和 $\mathbf{X}_j = \{x_{j,1}, x_{j,2}, x_{j,3}\} = \{1, 3, 5\}$。因此，通過公式(18)和公式(21)計算可得 $\mathbf{C}_i = \{c_{i,1}, c_{i,2}, c_{i,3}\} = \{6561, 13122, 19683\}$和 $\mathbf{C}_j = \{c_{j,1}, c_{j,2}, c_{j,3}\} = \{6561, 19683, 32805\}$，後續將以此為例說明基於同態加密的各個相似度計算方法。

為計算軟餘弦相似度，需先計算各個特徵之間的相似度。例如，$w_{k,l}$ 代表第 $k$ 個特徵和第 $l$ 個特徵的餘弦相似度，即計算 $\{x_{i,k}, x_{j,k}\}$ 和 $\{x_{i,l}, x_{j,l}\}$ 的餘弦相似度。因此，在此例中，向量 $\mathbf{X}_i$ 和向量 $\mathbf{X}_j$ 各個特徵之間的相似度為 $\{w_{1,1}, w_{1,2}, w_{1,3}, w_{2,1}, w_{2,2}, w_{2,3}, w_{3,1}, w_{3,2}, w_{3,3}\} \approx \{1, 0.98058, 0.97014, 0.98058, 1, 0.99887, 0.97014, 0.99887, 1\}$。

### 3.3.1 基於同態加密的餘弦相似度

本節運用公式(23)和公式(24)分別計算基於同態加密的餘弦相似度和餘弦距離，如公式(32)和公式(33)所示。通過計算實例結果可以驗證在密文的情況下相乘和加總後可以得到與明文計算後一致的結果：$s_{c,i,j} \approx 0.99386$、$d_{c,i,j} \approx 0.00614$。由此證明基於同態加密的餘弦相似度計算方法的可行性。

$$\frac{Dec(C_i \cdot C_j)}{Dec(\|C_i\|\|C_j\|)}$$

$$= \frac{\left[\left(\frac{6561}{9^5}\right)\left(\frac{6561}{9^5}\right) + \left(\frac{13122}{9^5}\right)\left(\frac{19683}{9^5}\right) + \left(\frac{19683}{9^5}\right)\left(\frac{32805}{9^5}\right)\right](\mod p)}{\left[\sqrt{\left(\frac{6561}{9^5}\right)^2 + \left(\frac{13122}{9^5}\right)^2 + \left(\frac{19683}{9^5}\right)^2}\sqrt{\left(\frac{6561}{9^5}\right)^2 + \left(\frac{19683}{9^5}\right)^2 + \left(\frac{32805}{9^5}\right)^2}\right](\mod p)} \quad (32)$$

$$= \frac{(1)(1) + (2)(3) + (3)(5)}{\sqrt{(1)^2 + (2)^2 + (3)^2}\sqrt{(1)^2 + (3)^2 + (5)^2}} = \frac{X_i \cdot X_j}{\|X_i\|\|X_j\|} = s_{c,i,j} \approx 0.99386$$

$$1 - \frac{Dec(C_i \cdot C_j)}{Dec(\|C_i\|\|C_j\|)} = 1 - \frac{X_i \cdot X_j}{\|X_i\|\|X_j\|} = 1 - s_{c,i,j} = d_{c,i,j} \approx 1 - 0.99386 = 0.00614 \quad (33)$$

$$1 - \frac{2 \times \arccos\left(\frac{Dec(C_i \cdot C_j)}{Dec(\|C_i\|\|C_j\|)}\right)}{\pi} = 1 - \frac{2 \times \arccos\left(\frac{X_i \cdot X_j}{\|X_i\|\|X_j\|}\right)}{\pi}$$

$$= 1 - \frac{2 \times \arccos(0.99386)}{\pi} = s_{a,i,j} \approx 0.92941 \quad (34)$$

$$\frac{2 \times \arccos\left(\frac{Dec(C_i \cdot C_j)}{Dec(\|C_i\|\|C_j\|)}\right)}{\pi} = \frac{2 \times \arccos(0.99386)}{\pi} = d_{a,i,j} \approx 0.07059 \quad (35)$$

$$\frac{Dec(C_i \cdot C_j)}{Dec(\|C_i\|^2) + Dec(\|C_j\|^2) - Dec(C_i \cdot C_j)}$$

$$= \left\{\left[\begin{array}{l}\left(\frac{6561}{9^5}\right)\left(\frac{6561}{9^5}\right) + \\ \left(\frac{13122}{9^5}\right)\left(\frac{19683}{9^5}\right) + \\ \left(\frac{19683}{9^5}\right)\left(\frac{32805}{9^5}\right)\end{array}\right](\mod p)\right\} \Bigg/ \left\{\begin{array}{l}\left[\left(\left(\frac{6561}{9^5}\right)^2 + \left(\frac{13122}{9^5}\right)^2 + \left(\frac{19683}{9^5}\right)^2\right)(\mod p)\right] + \\ \left[\left(\left(\frac{6561}{9^5}\right)^2 + \left(\frac{19683}{9^5}\right)^2 + \left(\frac{32805}{9^5}\right)^2\right)(\mod p)\right] - \\ \left[\begin{array}{l}\left(\frac{6561}{9^5}\right)\left(\frac{6561}{9^5}\right) + \\ \left(\frac{13122}{9^5}\right)\left(\frac{19683}{9^5}\right) + \\ \left(\frac{19683}{9^5}\right)\left(\frac{32805}{9^5}\right)\end{array}\right](\mod p)\end{array}\right\} \quad (36)$$

$$= \{[(1)(1) + (2)(3) + (3)(5)](\mod p)\} \Bigg/ \left\{\begin{array}{l}[((1)^2 + (2)^2 + (3)^2)(\mod p)] + \\ [((1)^2 + (3)^2 + (5)^2)(\mod p)] - \\ [(1)(1) + (2)(3) + (3)(5)](\mod p)\end{array}\right\}$$

$$= \frac{X_i \cdot X_j}{\|X_i\|^2 + \|X_j\|^2 - X_i \cdot X_j} = s_{t,i,j} \approx 0.81482$$

### 3.3.2 基於同態加密的角度相似度

本節運用公式(25)和公式(26)分別計算基於同態加密的角度相似度和角度距離，如公式(34)和公式(35)所示。通過計算實例結果可以驗證在密文的情況下相乘和加總後可以得到與明文計算後一致的結果：$s_{a,i,j} \approx 0.92941$、$d_{a,i,j} \approx 0.07059$。由此證

$$1 - \frac{Dec(C_i \cdot C_j)}{Dec(\|C_i\|^2) + Dec(\|C_j\|^2) - Dec(C_i \cdot C_j)} = 1 - \frac{X_i \cdot X_j}{\|X_i\|^2 + \|X_j\|^2 - X_i \cdot X_j} \quad (37)$$

$$= d_{t,i,j} \approx 0.18518$$

明基於同態加密的角度相似度計算方法的可行性。

### 3.3.3 基於同態加密的 Tanimoto 相似度

本節運用公式(27)和公式(28)分別計算基於同態加密的 Tanimoto 相似度和 Tanimoto 距離，如公式(36)和公式(37)所示。通過計算實例結果可以驗證在密文的情況下相乘和加總後可以得到與明文計算後一致的結果：$s_{t,i,j} \approx 0.81482$、$d_{t,i,j} \approx 0.18518$。由此證明基於同態加密的 Tanimoto 相似度計算方法的可行性。

### 3.3.4 基於同態加密的軟餘弦相似度

本節運用公式(30)和公式(31)分別計算基於同態加密的軟餘弦相似度和餘弦距離，如公式(38)和公式(39)所示。通過計算實例結果可以驗證在密文的情況下相乘和加總後可以得到與明文計算後一致的結果：$s_{s,i,j} \approx 0.99991$、$d_{s,i,j} \approx 0.00009$。由此證明基於同態加密的軟餘弦相似度計算方法的可行性。

## 4. 實證分析與討論

本節為實際驗證本研究提出的基於同態加密的各種相似度計算方法之效率，在實驗環境中採用一台 Windows 10 企業版的電腦執行演算法，計算加密時間(如 4.1 節)、基於同態加密的相似度計算時間(如 4.2 節)。其中，實驗使用的軟硬體詳細規格是 CPU Intel(R) Core(TM) i7-10510U、記憶體 8 GB、OpenJDK 18.0.2.1、以及函式庫 Bouncy Castle Release 1.72。

除此之外，本研究採用美國國家標準暨技術研究院所規範的安全強度標準來驗證不同安全強度下的效率，如表 1 所示[4]。其中，安全強度分共可為五個等級，並且取得每個等級對應 RSA 密碼學的金鑰長度作為私鑰長度參數來實作。

表 1 安全強度與 RSA 密碼學參數

| 安全強度 | RSA 金鑰長度(單位：bit) |
|---|---|
| 80 | 1024 |
| 112 | 2048 |
| 128 | 3072 |
| 192 | 7680 |
| 256 | 15360 |

### 4.1 加密時間

本節主要驗證不同安全強度下，採用本研究演算法所需的加密時間，即計算公式(18)所需時間。其中，在實驗環境中將隨機產生兩組向量各 1,000 個維度(即 $n = 1000$)，並對每一筆資料計算其密文。表 2 為對兩組向量各別加密的總時間長度，時間單位為毫秒。由實驗結果顯示，由於演算法採用金鑰作為指數項，所以隨著安全強度的增加，加密時間將快速增加。

表 2 加密時間(單位：毫秒)

| 安全強度 | 加密時間 |
|---|---|
| 80 | 7.046 |
| 112 | 10.196 |
| 128 | 44.776 |
| 192 | 249.014 |
| 256 | 1420.920 |

### 4.2 相似度計算時間

本節主要驗證不同安全強度下，採用本研究演算法所需的相似度計算時間。其中，基於同態加密的餘弦相似度計算方法採用公式(23)計算、基於同態加密的角度相似度計算方法採用公式(25)計算、基於同態加密的 Tanimoto 相似度計算方法採用公式(27)計算、基於同態加密的軟餘弦相似度計算方法採用公式(30)計算。其中，在實驗環境中已儲存在 4.1 節計算的密文，兩組向量各 1,000 個維度。

$$= \cfrac{\left\{\left[\begin{array}{c}(1)\left(\frac{6561}{9^5}\right)\left(\frac{6561}{9^5}\right)+\\(0.98058)\left(\frac{6561}{9^5}\right)\left(\frac{19683}{9^5}\right)+\\(0.97014)\left(\frac{6561}{9^5}\right)\left(\frac{32805}{9^5}\right)+\\(0.98058)\left(\frac{13122}{9^5}\right)\left(\frac{6561}{9^5}\right)+\\(1)\left(\frac{13122}{9^5}\right)\left(\frac{19683}{9^5}\right)+\\(0.99887)\left(\frac{13122}{9^5}\right)\left(\frac{32805}{9^5}\right)+\\(0.97014)\left(\frac{19683}{9^5}\right)\left(\frac{6561}{9^5}\right)+\\(0.99887)\left(\frac{19683}{9^5}\right)\left(\frac{19683}{9^5}\right)+\\(1)\left(\frac{19683}{9^5}\right)\left(\frac{32805}{9^5}\right)\end{array}\right](\bmod\ p)\right\}}{\left\{\sqrt{\left[\begin{array}{c}(1)\left(\frac{6561}{9^5}\right)\left(\frac{6561}{9^5}\right)+\\(0.98058)\left(\frac{6561}{9^5}\right)\left(\frac{13122}{9^5}\right)+\\(0.97014)\left(\frac{6561}{9^5}\right)\left(\frac{19683}{9^5}\right)+\\(0.98058)\left(\frac{13122}{9^5}\right)\left(\frac{6561}{9^5}\right)+\\(1)\left(\frac{13122}{9^5}\right)\left(\frac{13122}{9^5}\right)+\\(0.99887)\left(\frac{13122}{9^5}\right)\left(\frac{19683}{9^5}\right)+\\(0.97014)\left(\frac{19683}{9^5}\right)\left(\frac{6561}{9^5}\right)+\\(0.99887)\left(\frac{19683}{9^5}\right)\left(\frac{13122}{9^5}\right)+\\(1)\left(\frac{19683}{9^5}\right)\left(\frac{19683}{9^5}\right)\end{array}\right]}\sqrt{\left[\begin{array}{c}(1)\left(\frac{6561}{9^5}\right)\left(\frac{6561}{9^5}\right)+\\(0.98058)\left(\frac{6561}{9^5}\right)\left(\frac{19683}{9^5}\right)+\\(0.97014)\left(\frac{6561}{9^5}\right)\left(\frac{32805}{9^5}\right)+\\(0.98058)\left(\frac{19683}{9^5}\right)\left(\frac{6561}{9^5}\right)+\\(1)\left(\frac{19683}{9^5}\right)\left(\frac{19683}{9^5}\right)+\\(0.99887)\left(\frac{19683}{9^5}\right)\left(\frac{32805}{9^5}\right)+\\(0.97014)\left(\frac{32805}{9^5}\right)\left(\frac{6561}{9^5}\right)+\\(0.99887)\left(\frac{32805}{9^5}\right)\left(\frac{19683}{9^5}\right)+\\(1)\left(\frac{19683}{9^5}\right)\left(\frac{32805}{9^5}\right)\end{array}\right]}(\bmod\ p)\right\}}$$

$$= \frac{\sum_{k=1}^n \sum_{l=1}^n w_{k,l} x_{i,k} x_{j,l}}{\sqrt{\sum_{k=1}^n \sum_{l=1}^n w_{k,l} x_{i,k} x_{i,l}} \sqrt{\sum_{k=1}^n \sum_{l=1}^n w_{k,l} x_{j,k} x_{j,l}}} = s_{s,i,j} \approx 0.99991 \quad (38)$$

$$1 - \frac{Dec\left(\sum_{k=1}^n \sum_{l=1}^n \omega_{k,l} c_{i,k} c_{j,l}\right)}{Dec\left(\sqrt{\sum_{k=1}^n \sum_{l=1}^n \omega_{k,l} c_{i,k} c_{i,l}} \sqrt{\sum_{k=1}^n \sum_{l=1}^n \omega_{k,l} c_{j,k} c_{j,l}}\right)} = 1 - s_{s,i,j} = d_{s,i,j} \approx 0.00009 \quad (39)$$

表 3 為對兩組向量各 1,000 個維度進行相似度計算的時間，時間單位為毫秒。由實驗結果顯示，隨著安全強度增加，相似度計算的時間也將增加。其中，由於餘弦相似度和角度相似度計算類似，所以計算時間無顯著差異。而 Tanimoto 相似度計算上不需開平方根，所以計算時間可以較少。軟餘弦相似度的計算時間會依維度呈指數成長，所以需要較計算時間。

表 3 相似度計算時間(單位：毫秒)

| 相似度 | 80 | 112 | 128 | 192 | 256 |
|---|---|---|---|---|---|
| 餘弦相似度 | 8.116 | 10.356 | 31.654 | 118.173 | 245.597 |
| 角度相似度 | 8.181 | 10.616 | 31.741 | 118.615 | 245.658 |
| Tanimoto 相似度 | 3.409 | 5.336 | 15.597 | 59.862 | 201.337 |
| 軟餘弦相似度 | 2190.102 | 5950.921 | 13942.654 | 87455.435 | 284729.025 |

## 5. 結論

本研究在 ElGamal 加解密方法的基礎上，提出基於同態加密的各種相似度計算方法，包含有基於同態加密的餘弦相似度計算方法、基於同態加密的角度相似度計算方法、基於同態加密的 Tanimoto 相似度計算方法、基於同態加密的軟餘弦相似度計算方法。在 3.2 節中提出數學模型證明，在 3.3 節提供計算實例說明。並且，在實驗中驗證在不同安全強度下，基於同態加密的各種相似度計算方法的效率，以及討論各個方法的效率差異。

## 參考文獻

1. L. Zhang, et al., "Homomorphic Encryption-based Privacy-preserving Federated Learning in IoT-enabled Healthcare System," *IEEE Transactions on Network Science and Engineering*, doi: 10.1109/TNSE.2022.3185327
2. T. Elgamal, "A public key cryptosystem and a signature scheme based on discrete logarithms," *IEEE Transactions on Information Theory*, vol. 31, no. 4, pp. 469-472, 1985.
3. G. Sidorov, et al., " Soft Similarity and Soft Cosine Measure: Similarity of Features in Vector Space Model," *Computación y Sistemas*, vol. 18, no. 3, pp. 491-504, 2014.
4. E. Barker, "Recommendation for Key Management: Part 1 – General," *NIST Special Publication*, NIST SP 800-57 Part 1 Rev. 5, 2020. doi: 10.6028/NIST.SP.800-57pt1r5

# Similarity Calculation Based on Homomorphic Encryption

Abel C. H. Chen[1*]
[1]Chunghwa Telecom Co., Ltd.
*chchen.scholar@gmail.com

**Abstract**

In recent years, although some homomorphic encryption algorithms have been proposed to provide additive homomorphic encryption and multiplicative homomorphic encryption. However, similarity measures are required for searches and queries under homomorphic encrypted ciphertexts. Therefore, this study considers the cosine similarity, angular similarity, Tanimoto similarity, and soft cosine similarity and combines homomorphic encryption algorithms for similarity calculation. This study proposes mathematical models to prove the proposed homomorphic encryption-based similarity calculation methods and gives practical cases to explain the proposed methods. In experiments, the performance of the proposed homomorphic encryption-based similarity calculation methods has been evaluated under different security strengths.

***Keywords***: *Homomorphic encryption, similarity, cryptography*